\documentclass[aip,amsmath,amssymb]{revtex4-1}
\usepackage{graphicx}

\usepackage{dcolumn}
\usepackage{bm}
\usepackage[utf8]{inputenc}
\usepackage[T1]{fontenc}
\usepackage{mathptmx}
\usepackage{etoolbox}

\begin{document}
\title{Molecular Dynamics Study of Rayleigh-Plateau Instability at Liquid-Liquid Interfaces}

\author{Shunta Kikuchi\thanks{skkt.9x18@keio.jp} and Hiroshi Watanabe\thanks{hwatanabe@appi.keio.ac.jp}}
\affiliation{
    Department of Applied Physics and Physico-Informatics, Faculty of Science and Technology, Keio University, Kanagawa 233-8522, Japan
}

\date{\today}

\begin{abstract}
    We investigated the Rayleigh-Plateau instability at the interface between two immiscible liquids of equal viscosity using molecular dynamics simulations. Two types of initial conditions were considered,  one with an imposed single-mode perturbation at the interface and the other without any imposed perturbation. Under the single-mode perturbation, the growth rate deviated from the theoretical prediction for small cylinder radii, but progressively approached and agreed with classical macroscopic theory as the radius increased. In contrast, for the unperturbed initial condition, we found a systematic relationship between the breakup time and the minimum radius, in which the power-law exponent increased with increasing radius. These results demonstrate that, even in extremely microscopic systems with cylinder radii on the order of only about fifteen atomic diameters, the growth of the instability can follow macroscopic theoretical predictions when appropriate conditions are imposed, and that the influence of thermal fluctuations on the breakup dynamics becomes increasingly significant as the system radius decreases.
\end{abstract}

\maketitle

\section{Introduction}
The Rayleigh-Plateau (RP) instability is a phenomenon in fluid dynamics where a cylindrical fluid becomes unstable and breaks up into smaller droplets\cite{plateau1873experimental,rayleigh1878instability,eggers2008physics}. Because RP instability plays essential roles in various technical fields, such as inkjet printing and circuit board manufacturing, understanding RP instability can improve these processes\cite{basaran2013nonstandard,lohse2022fundamental,day2015plateau}. In particular, the RP instability at liquid-liquid interfaces is essential for applications such as drug delivery via the injections and the formation of liposomes, suggesting significant potential for advances in pharmaceutical research\cite{stachowiak2009dynamic}. The RP instability has been extensively studied experimentally, theoretically, and numerically\cite{eggers2008physics,basaran2013nonstandard,lohse2022fundamental}. Rayleigh conducted a linear stability analysis for inviscid systems in the early theoretical study. Subsequently, Stone and Tomotika extended this study to viscous systems, deriving dispersion relations for growth rates in the absence of thermal fluctuations\cite{rayleigh1878instability,stone1996note,tomotika1935instability}. The validity of these theories has been examined by comparing the dispersion relations for growth rates, breakup times, droplet size distributions, and the power law behavior of the minimum radius\cite{petit2012break,geschiere2012slow}.

The particle-based numerical methods, such as dissipative particle dynamics and molecular dynamics (MD), allow us to investigate the behavior of the breakup on a molecular scale. These approaches can also quantify detailed physical properties inaccessible to experiments, enabling direct comparisons with theoretical predictions. The numerical studies have examined the cylindrical geometry near breakup, the size distribution of droplets post-breakup, and the influence of surfactants on destabilization\cite{gopan2014rayleigh,tiwari2008dissipative,carnevale2024surfactant}. In particular, thermal fluctuations have been found to affect the pre-breakup interface shape, indicating that fluid equations incorporating thermal fluctuation terms become dominant on microscopic scales\cite{moseler2000formation,zhao2019revisiting}. The correspondence between theoretical predictions and numerical simulations has also been discussed in previous studies. Eggers et al.~explained, using a path-integral approach, that thermal fluctuations reduce the breakup exponent and lead to more abrupt rupture dynamics\cite{eggers2002dynamics}, while Gopan et al.~evaluated the validity of this theory through numerical simulations of gas-liquid systems\cite{gopan2014rayleigh}.

Previous particle-based simulations of the RP instability have mostly focused on gas-liquid interfaces. However, such systems require separate thermal equilibration of the gas and liquid phases before coupling them, and in addition to interfacial instability, the effects of evaporation must also be taken into account. To address this difficulty, we investigated the instability of the interface between two liquids with identical physical properties by MD simulations. We first performed simulations of single-component fluids to prepare the system with the liquid-liquid interface and obtained the equilibrium state. We then generated liquid-liquid interfaces from the resulting single-component liquids by setting different types of atoms inside and outside the cylinder shape. In this way, we achieved conditions in which properties are identical inside and outside the interface, satisfying the conditions required by Stone's theory. Therefore, we can directly compare the simulation results to the theory. Furthermore, the interface between liquids that do not mix is more stable than the gas-liquid interface and is not affected by anything other than RP instability.

Molecular dynamics enables phenomena to be resolved from the atomic scale and thus allows direct investigation of how microscopic structures influence macroscopic behavior. However, because the system size is inherently small due to the limitation of the computational resources, the results are affected by the discreteness of particles. In the RP instability, thermal fluctuations play an important role, and in addition, the effects of particle discreteness arising from the finite system size cannot be neglected. To isolate the influence of particle discreteness, we imposed initial interfacial perturbations with amplitudes sufficiently larger than those induced by thermal fluctuations, thereby suppressing thermal effects. We found that the growth rate deviates from Stone's theoretical prediction at small cylinder radii, whereas good agreement with the theory is observed at larger radii. The fact that good agreement with macroscopic theory is obtained even for extremely small systems with cylinder radii on the order of only about fifteen atomic diameters suggests that the influence of particle discreteness on the RP instability is very small.

To quantitatively investigate the influence of thermal fluctuations on the RP instability, we examined the relationship between the breakup time and the minimum radius under unperturbed conditions. It is known that the minimum radius of a liquid column just before breakup is proportional to a power of the remaining time to breakup, and that the corresponding exponent is affected by thermal fluctuations~\cite{eggers1993universal,moseler2000formation, eggers2002dynamics}. By analyzing the breakup dynamics of liquid cylinder with various radii, we investigated the radius dependence of this exponent. We found that the value of the breakup exponent increases as the cylinder radius becomes larger. This result is consistent with the theoretical prediction that the breakup exponent approaches $1$ in the absence of thermal fluctuations.

The rest of the paper is organized as follows. We describe the method in section \ref{sec:method}. The results are described in Sec.~\ref{sec:results}. Section~\ref{sec:summary} is devoted to the summary and discussion.

\section{Method} \label{sec:method}

\subsection{Linear stability analysis}

Stone and Brenner described the linear stability analysis of the RP instability for fluids with equal viscosities\cite{stone1996note}. When thermal fluctuations are not considered, previous studies have shown that the dimensionless growth rate $\tilde{\omega}(\tilde{k})$ can be described as the dimensionless wavenumber $\tilde{k} = r_0k$:
\begin{equation}
    \tilde{\omega}(\tilde{k}) = \frac{\mu r_0}{\gamma} \omega(\tilde{k}) =  (1- \tilde{k}^2) \left[ I_1(\tilde{k})K_1(\tilde{k}) + \frac{\tilde{k}}{2}(I_1(\tilde{k})K_0(\tilde{k})-I_0(\tilde{k})K_1(\tilde{k})) \right],
    \label{eq:theory_normalized}
\end{equation}
where $I$ and $K$ denote the modified Bessel functions of the first and second kinds, respectively. We determined the growth rate $\omega (k)$ using MD simulations.

Then we determined dimensionless growth rate $\tilde{\omega}(\tilde{k})$ by the interfacial tension $\gamma$ and viscosity $\mu$, which we determined separately.

\subsection{Molecular dynamics simulations}

We adopted the Lennard-Jones(LJ) potential for atomic interactions. The LJ potential is defined as
\begin{equation}
    \phi(r) = \left\{
    \begin{array}{cc}
        4 \varepsilon \left[ \left( \displaystyle \frac{\sigma}{r}\right)^{12} - \left( \displaystyle \frac{\sigma}{r}\right)^{6} \right] & (r \geq r_\mathrm{c}), \\
        0                                                                                                                                 & (r \leq r_\mathrm{c}),
    \end{array}
    \right.
    \label{eq:lj}
\end{equation}
where $\sigma$ is the atomic diameter, $\varepsilon$ is the interaction energy, $r$ is the interatomic distance, and $r_\mathrm{c}$ is the cutoff distance\cite{lennard1931cohesion,weeks1971role}, respectively. In the following, we adopted the reduced unit such that the $\sigma, \varepsilon$, and $k_\mathrm{B}$ are unity. We set the cutoff $r_\mathrm{c}=3.0$ for interactions between atoms of the same type to represent attractive interactions. On the other hand, we set $r_\mathrm{c}=2^{1/6} \approx 1.12246$ for atoms with different types, which is the Weeks-Chandler-Andersen potential, to represent repulsive interactions. We used the velocity-Verlet algorithm with a time step of 0.005\cite{verlet1967computer}. The simulations were performed in the NVT ensemble. The size of the simulation box is $(L_x,L_y,L_z) = (100,100,1000)$. The total number of atoms is $8000000$. We defined the axial direction of the cylinder to be the $z$ direction. The temperature was maintained with Langevin thermostat\cite{paquet2015molecular,PhysRevB.17.1302}. Throughout this study, simulations were performed at a temperature of $0.8$. The damping parameter of the Langevin thermostat $\tau_\mathrm{damp}$ is set to be $50$. We performed the MD simulations using the Large-scale Atomic/Molecular Massively Parallel Simulator (LAMMPS)\cite{LAMMPS}.

We determined the interfacial tension to be $\gamma = 1.38$ ($r_0=5$), $\gamma = 1.54$ ($r_0=10$) and $\gamma = 1.60$ ($r_0=15$) from the Young-Laplace relation (see Appendix.~\ref{sec:appendix_itf}). The viscosity was determined to be $\mu = 2.47$ using the Kubo-Green method.

\subsection{Initial conditions}

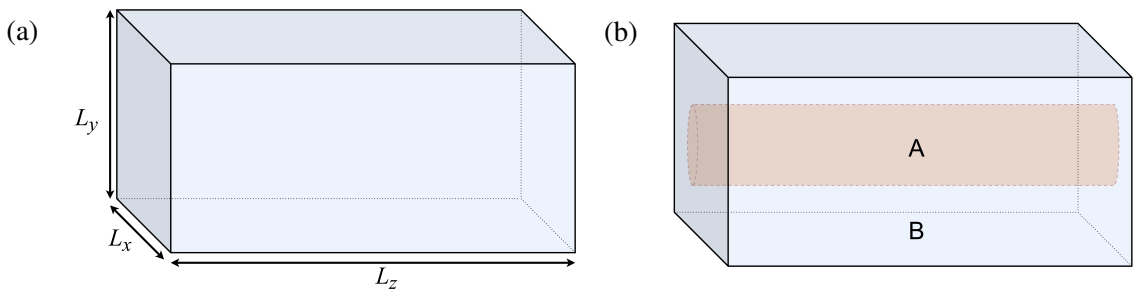
\begin{figure}[htbp]
    \begin{picture}(10,100)
        \put(0,100){(a)}
    \end{picture}
    \includegraphics[width = 0.45\textwidth]{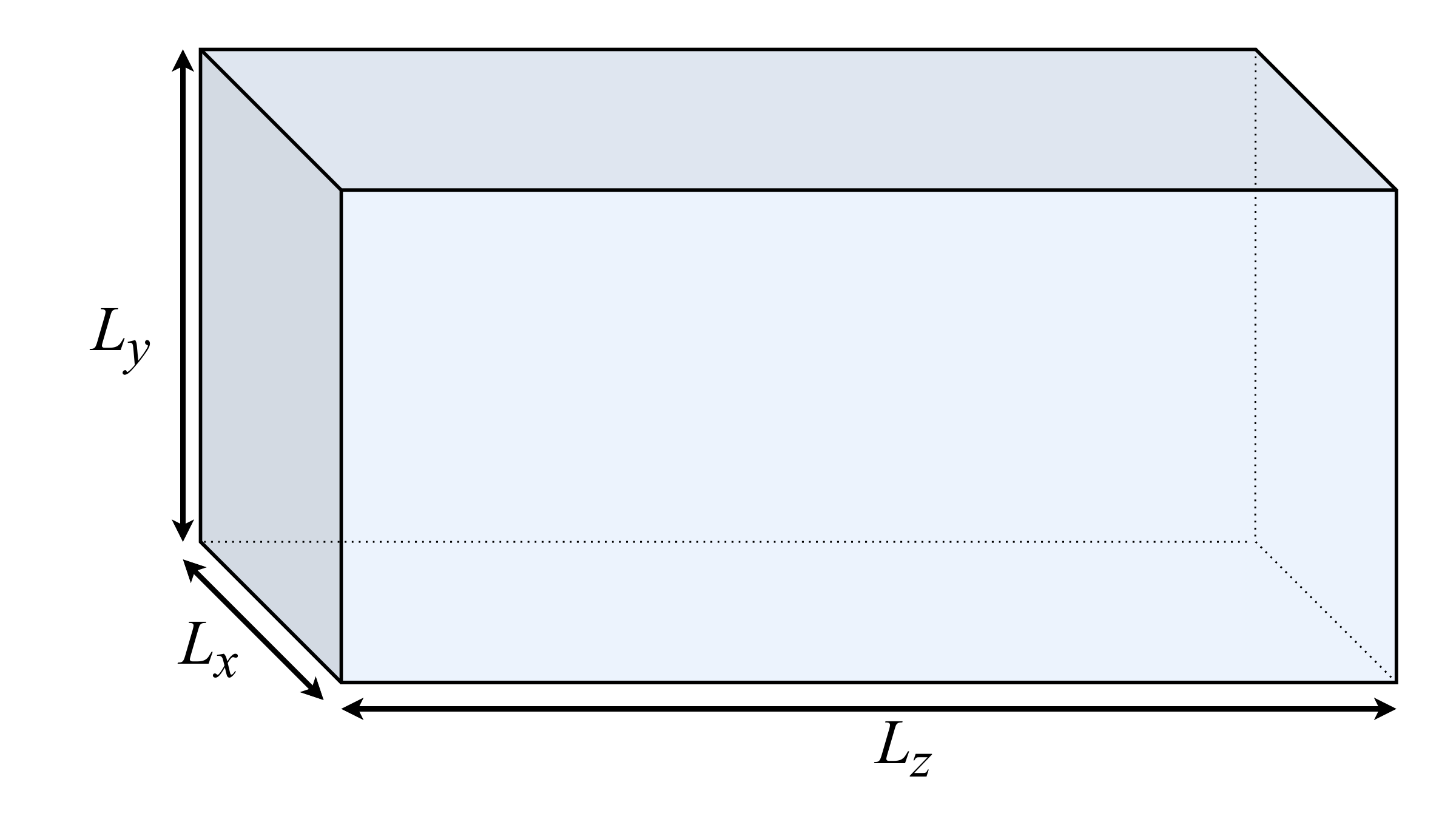}
    \begin{picture}(10,100)
        \put(0,100){(b)}
    \end{picture}
    \includegraphics[width=0.45\textwidth]{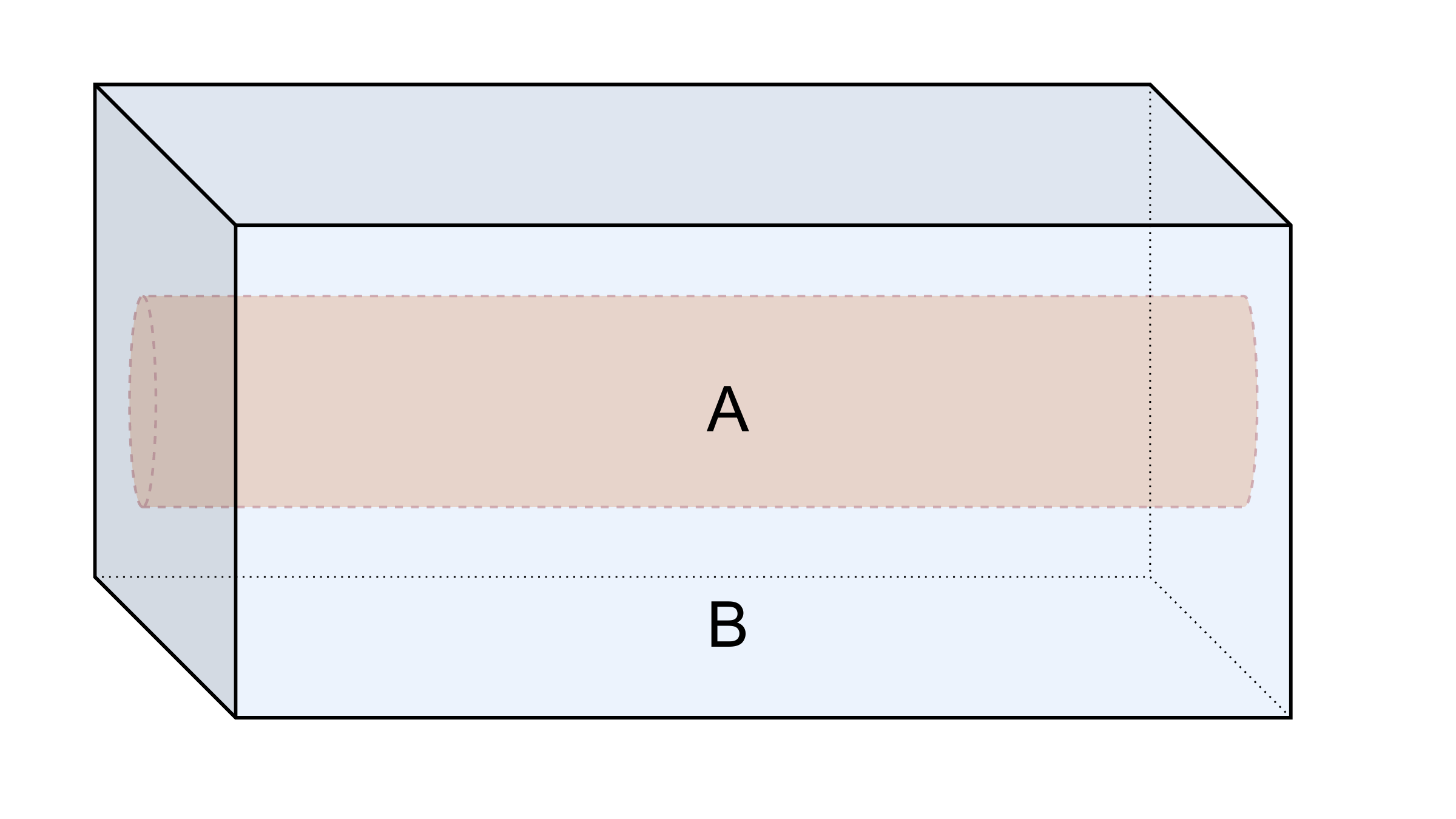}
    \caption{The generation of the initial configuration. (a) We first thermalized a single-component system. (b) After relaxation of the single-component system, we changed the labels of atoms in a cylinder of a specified radius.}
    \label{fig:rp_initial}
\end{figure}

In order to study RP instability, it is important to create initial conditions. The procedure for generating initial conditions is shown in Fig.~\ref{fig:rp_initial}. In order to create a stable interface, we first thermalized a single-component system (Fig.~\ref{fig:rp_initial}~(a)) and then changed the type of atoms inside the cylinder (Fig.~\ref{fig:rp_initial}~(b)). Then we have the initial conditions with a cylindrical liquid-liquid interface.

We investigated the two kinds of initial configuration, the interface with and without perturbation. We adopted the perturbed condition to compare the results with Stone's theory and the unperturbed condition to investigate the effect of thermal fluctuations.

\begin{figure}[htbp]
    \centering
    \begin{picture}(1000,10)
        \put(0,5){\large (a)}
    \end{picture}
    \includegraphics[scale = 0.5]{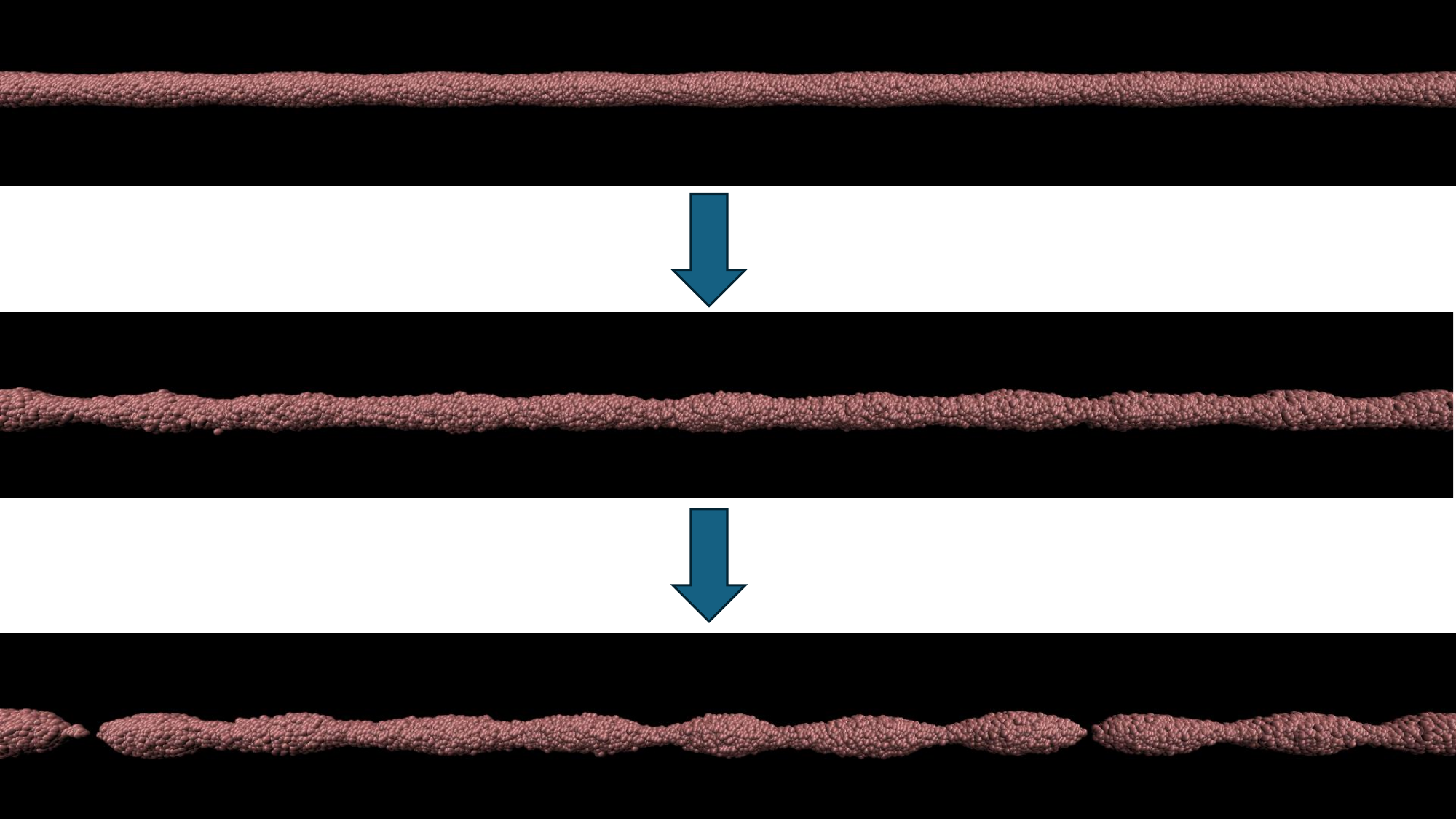}
    \begin{picture}(1000,10)
        \put(0,5){\large (b)}
    \end{picture}
    \includegraphics[scale = 0.5]{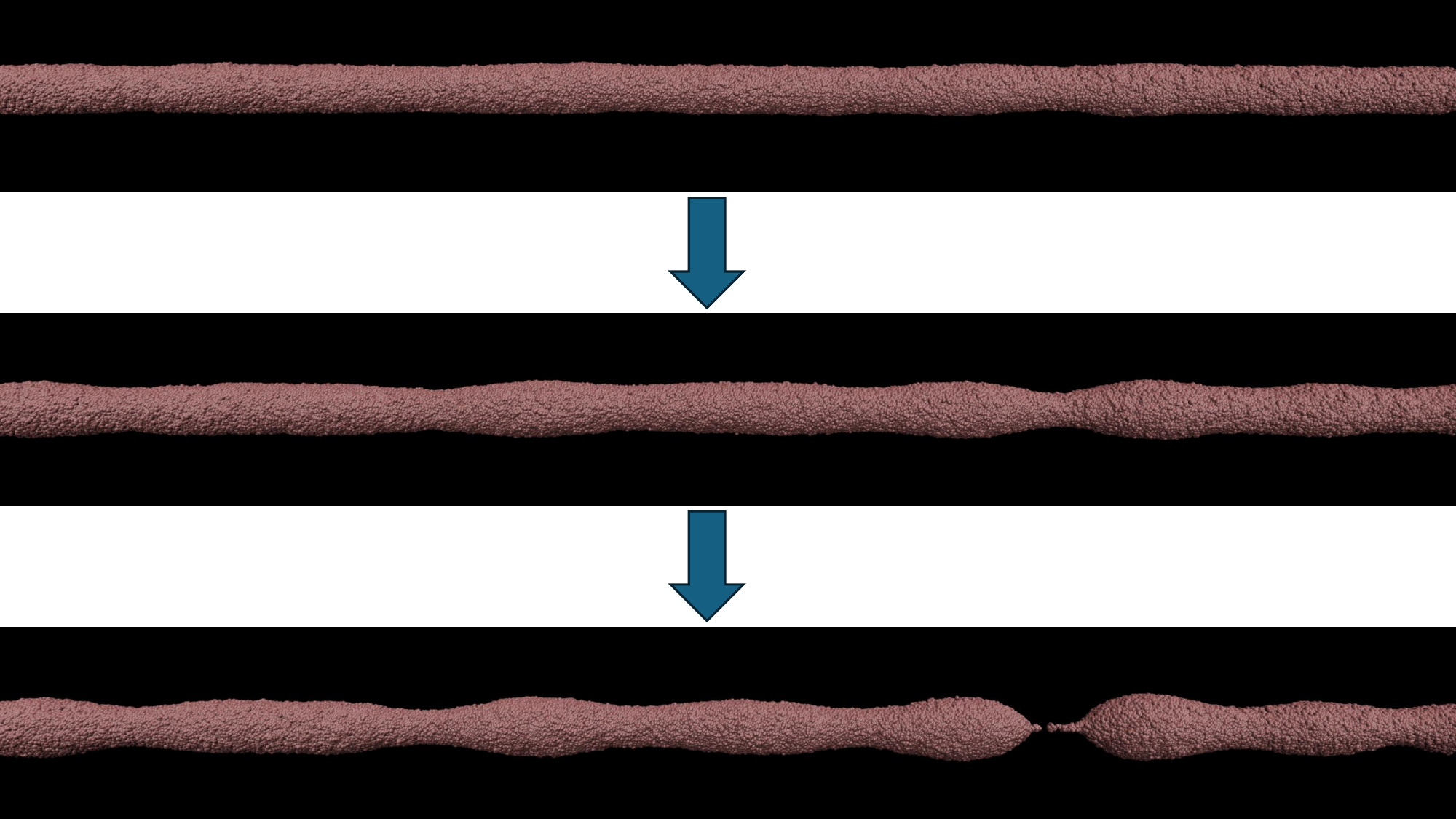}
    \caption{The snapshots of the simulations. (a) The breakup process of the cylinder with the imposed perturbation. The cylinder breaks up at evenly spaced intervals due to the imposed perturbation. (b) The breakup process of the cylinder without an imposed perturbation. The constriction occurs irregularly due to no perturbation.}
    \label{fig:rp_simulation}
\end{figure}
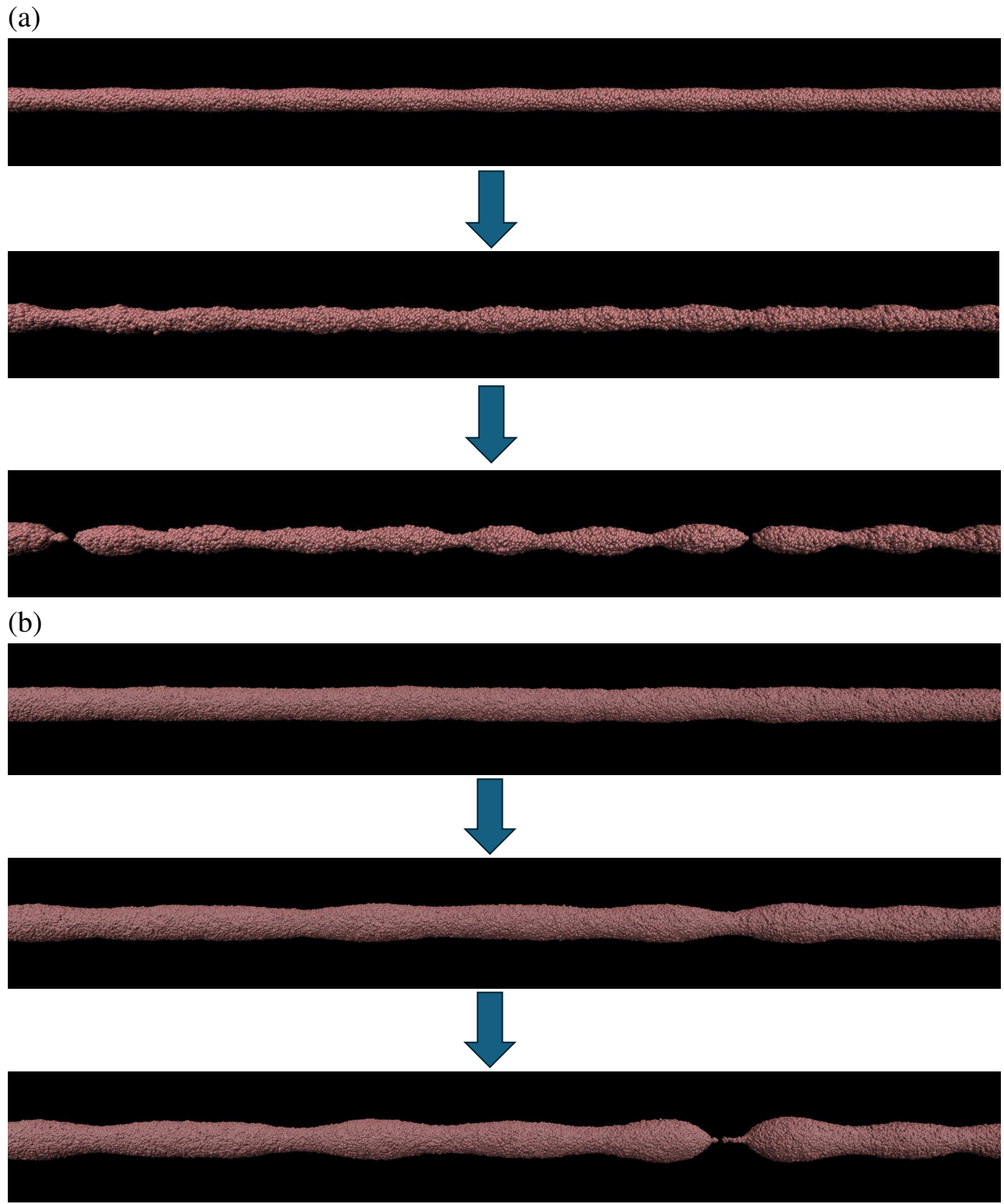

The typical time evolutions of the simulations are shown in Fig.~\ref{fig:rp_simulation}. Initially, the cylinder maintained its shape, but it was deformed and eventually broken. The grid was divided into 1000 segments along the $z$-direction. Additionally, the breakup time was defined as the first time when a grid segment with zero density appeared.

\subsection{Perturbed configuration}

\begin{figure}[htbp]
    \centering
    \begin{picture}(1000,10)
        \put(0,5){\large (a)}
    \end{picture}
    \includegraphics[width = 0.95\textwidth]{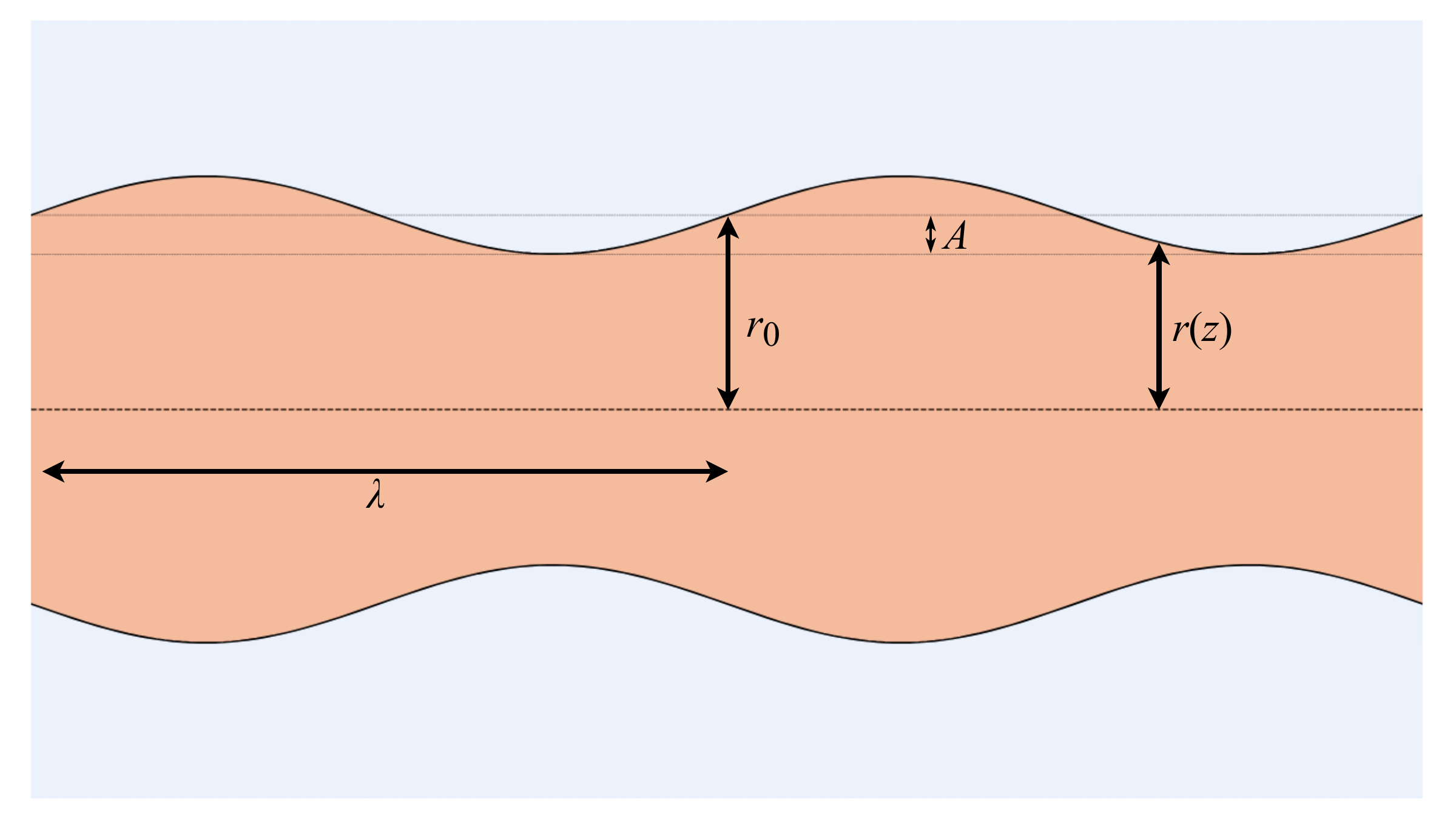}
    \begin{picture}(1000,10)
        \put(0,5){\large (b)}
    \end{picture}
    \includegraphics[width = 0.95\textwidth]{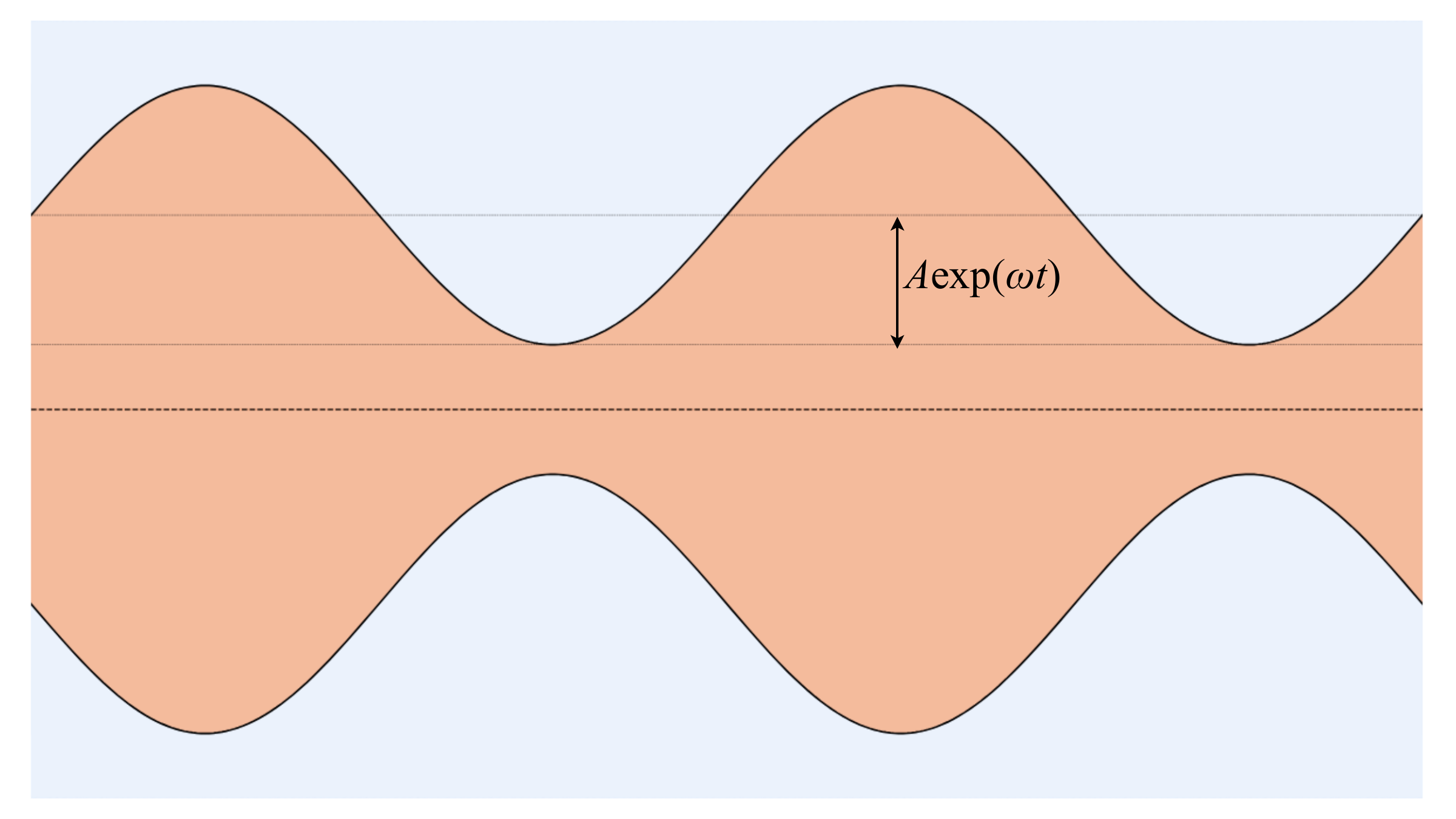}
    \caption{The schematic view of the cylinder with the imposed perturbation. (a) The schematic view of the cylinder at $t=0$ with the radius $r_0$, the amplitude $A$ and the wavelength of the perturbation $\lambda$. (b) The schematic view of the cylinder at time $t$. If only the perturbation with wavelength $\lambda$ grows, the amplitude evolves as $A\exp (\omega t)$ over time.}
    \label{fig:rp_param_perdu}
\end{figure}

The configuration with perturbation is shown in Fig.~\ref{fig:rp_param_perdu}~(a). The perturbation was imposed as the sinusoidal modulation with the wavenumber $k$ as
\begin{equation}
    r(z,0) = r_0 + A \cos(kz),
    \label{eq:cylinder}
\end{equation}
where $r_0$ is the radius of the cylinder and $A$ is the amplitude of the perturbation. We set $A = r_0/10$ throughout this study. Note that the wavenumber satisfies $kL_z=2n\pi$ due to the periodic boundary condition with an integer $n$. Assuming that only the perturbation mode grows (Fig.~\ref{fig:rp_param_perdu} (b)), we have the time evoluation of the radius of the cylinder at the position $z$ and time $t$ as
\begin{equation}
    r(z,t) = r_0 + A\cos(kz)\exp(\omega t),
    \label{eq:growth}
\end{equation}
where $\omega$ is the growth rate.

We determined growth rates by two methods, from a breakup time and Fourier transform. As the perturbation grows, eventually there will be a place where the radius becomes zero, which is the breakup of the liquid cylinder. Suppose $t_c$ is the breakup time. Assuming that the growth rate does not change until the breakup, we obtain the following equation
\begin{equation}
    r_0 = A \exp(\omega t_c).
\end{equation}
Solving the above equation for omega yields
\begin{equation}
    \omega = \frac{\ln(r_0/A)}{t_c}.
    \label{eq:RP_breakup_omega}
\end{equation}
Repeating simulations for various wavenumbers $k$, we determined the dispersion relation $\omega(k)$ from the observed breakup time $t_c$ using Eq.~(\ref{eq:RP_breakup_omega}). While this method of measuring breakup time has been widely employed in macroscopic experiments\cite{kalaaji2003breakup}, there are few examples of its use in microscopic simulations.

We also determined the dispersion relation from the Fourier transform. The time evolution of the amplitude of the perturbation $A(t)$ is determined by Fourier transforming the radius of the cylinder as follows.

\begin{equation}
    A(t) \propto \int r(z, t) \cos kz dz.
\end{equation}
From Eq.~(\ref{eq:growth}), the amplitude $A(t)$ is expressed to be
\begin{equation}
    A(t) = A\exp (\omega t). \label{eq:omega_from_exp}
\end{equation}
Therefore, the growth rate $\omega$ can be determined from the time evolution of the amplitude $A(t)$. By repeating the above procedure at various wavenumbers $k$, the dispersion relation $\omega(k)$ can be obtained. Since this method first imposes a perturbation with the specific wavenumber and then follows the time evolution of its amplitude, the dispersion relation can be obtained accurately even in negative growth rates, i.e., in stable modes.

\subsection{Unperturbed configuration}

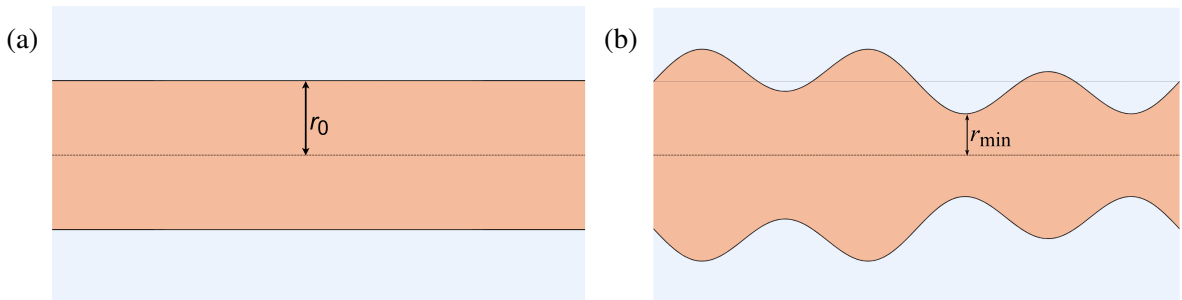
\begin{figure}[htbp]
    \centering
    \begin{picture}(10,100)
        \put(0,100){(a)}
    \end{picture}
    \includegraphics[width = 0.45\textwidth]{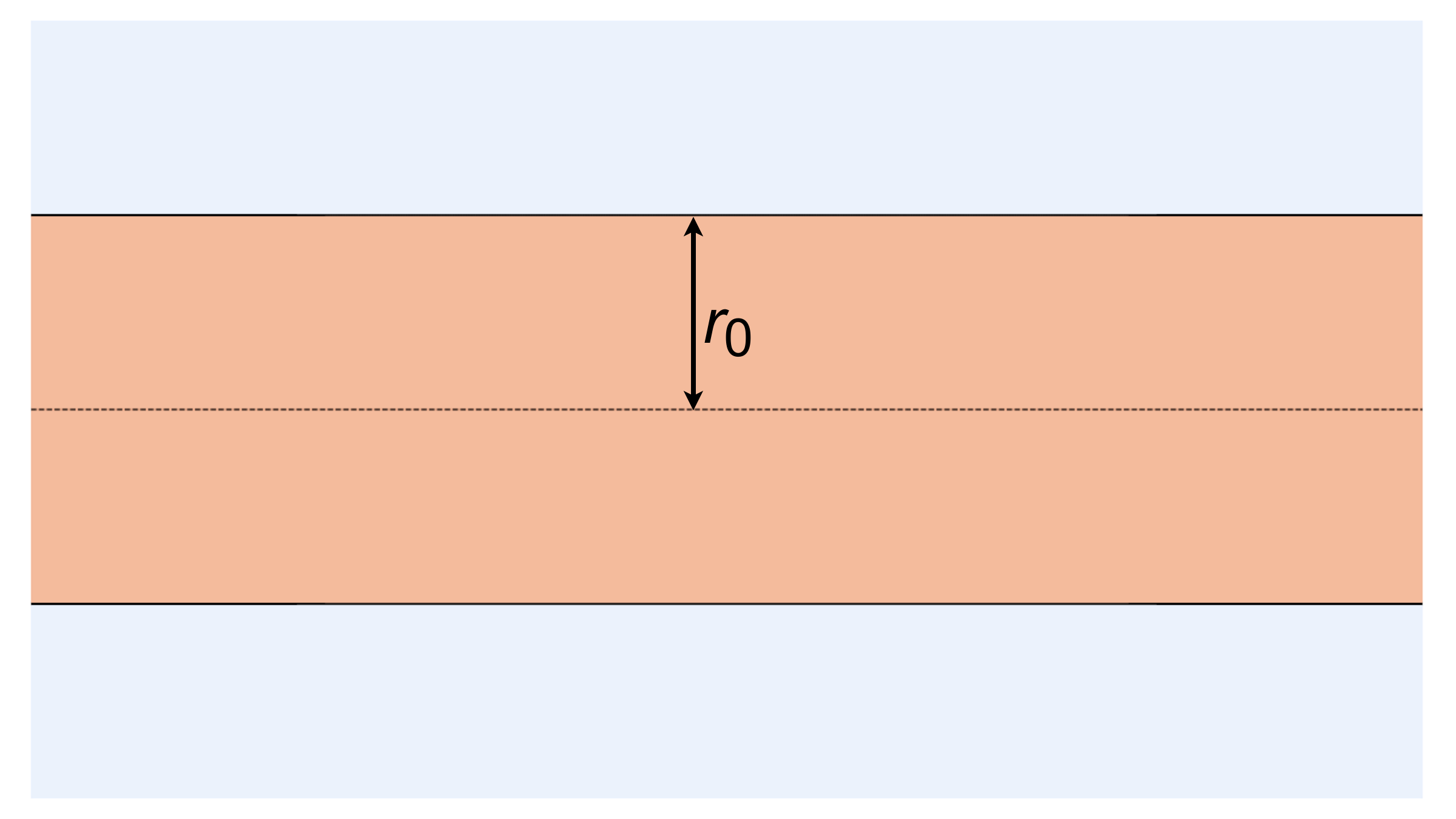}
    \begin{picture}(10,100)
        \put(0,100){(b)}
    \end{picture}
    \includegraphics[width = 0.45\textwidth]{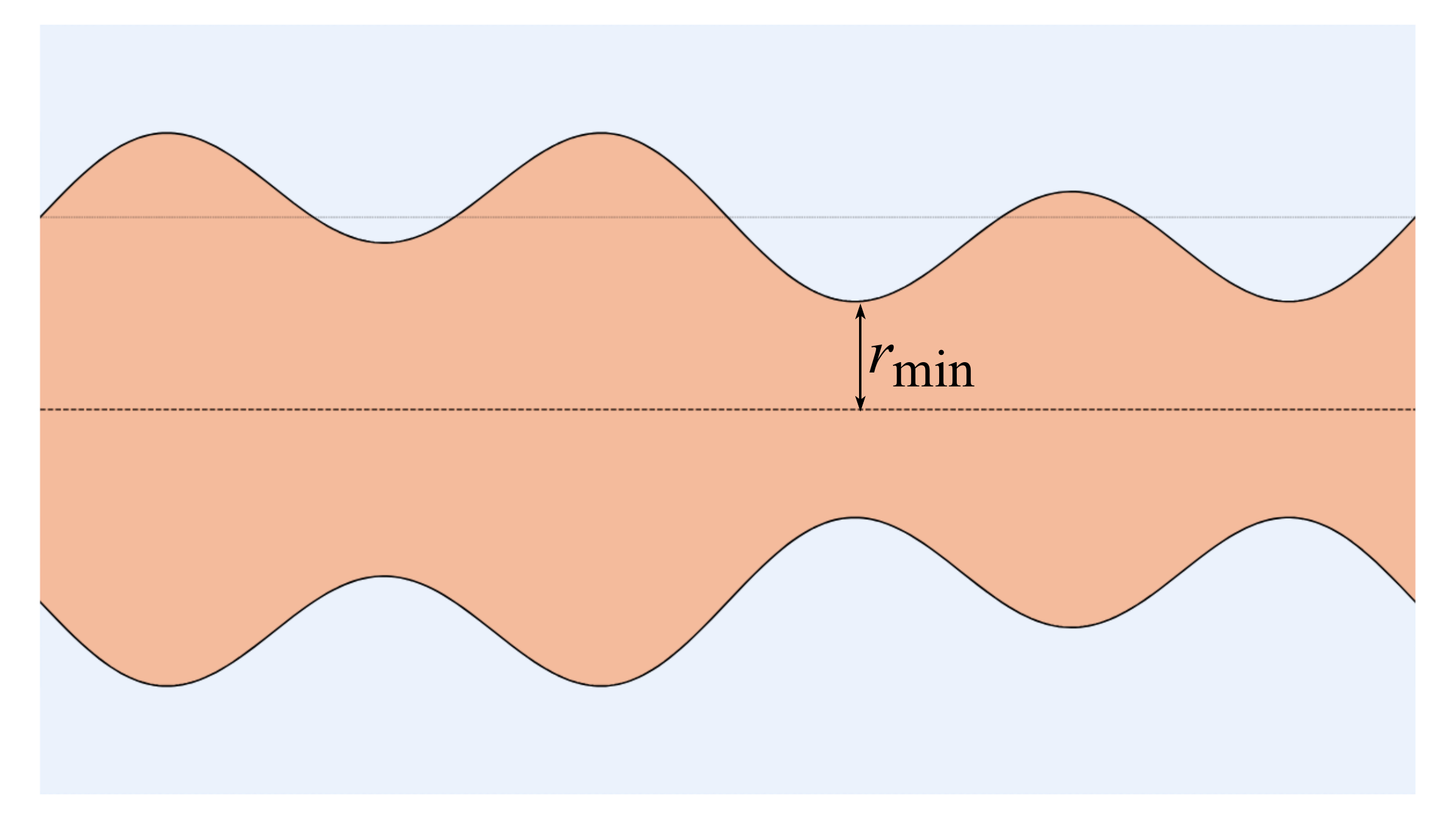}
    \caption{The schematic diagram of the cylinder without an imposed perturbation. (a) The schematic view of the cylinder with the initial radius $r_0$. (b) The schematic view of the cylinder at time $t$, when the minimum radius $r_{\min} (t)$ was measured in order to investigate the effects of the thermal fluctuations on the cylinder.}
    \label{fig:rp_param_nonperdu}
\end{figure}

For the cylinder without an imposed perturbation (Fig.~\ref{fig:rp_param_nonperdu} (a)), the effects of thermal fluctuations were analyzed through the deformation process. In this study, we measured the minimum radius $r_\mathrm{min} (t)$, which corresponds to the narrowest region of the cylinder (Fig.~\ref{fig:rp_param_nonperdu} (b)). The evolution of the minimum radius $r_\mathrm{min}$ near the breakup time follows:
\begin{equation}
    r_\mathrm{min} \propto (t_c-t)^\alpha.
\end{equation}
Here, $\alpha$ is an exponent that characterizes the effect of thermal fluctuations. Previous studies have shown that $\alpha = 1$ in the absence of thermal fluctuations, whereas $\alpha = 0.418$ when thermal fluctuations are present\cite{eggers1993universal,eggers2002dynamics}. In this study, the radius dependence of the exponent was determined, and its values were compared to investigate the influence of thermal fluctuations.

\section{Results}\label{sec:results}

\subsection{Perturbed Configuration}

To compare with the predictions of linear stability analysis, we determined the dispersion relation from the growth rate in a cylinder with imposed perturbation.

\begin{figure}[htbp]
    \centering
    \includegraphics[]{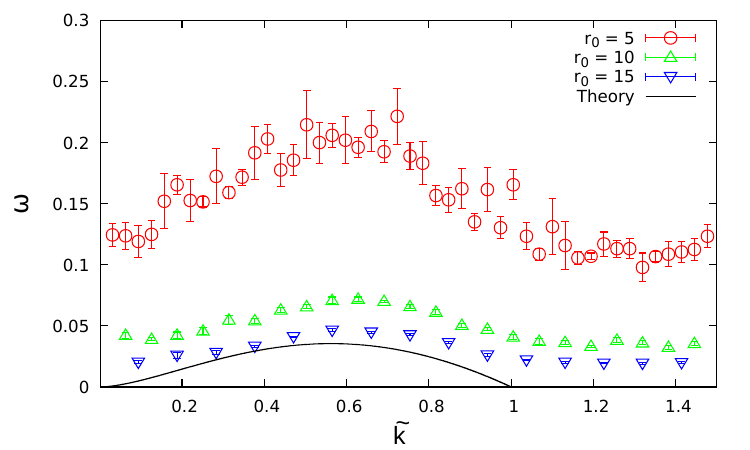}
    \caption{The dispersion relation calculated by the breakup time. The horizontal axis is the wavenumber, and the vertical axis is the growth rate. The growth rate does not become zero at $\tilde{k} = 0$ or $\tilde{k} \geq 1$.}
    \label{fig:RP_breakup_omega}
\end{figure}

First, we performed time evolution of the system from the initial condition with a specified wavenumber perturbation and observed the breakup time $t_c$, which is the time until the cylinder breaks. Then we determined the growth rate $\omega$ using the relation Eq.~(\ref{eq:RP_breakup_omega}). By repeating this procedure for each wavenumber, we determined the wavenumber dependence of the growth rate. The wavenumber dependence of the growth rate is shown in Fig.~\ref{fig:RP_breakup_omega}.

In this study, we performed five simulations for each parameter set and calculated the error bars from the results. As the initial radius of the cylinder increases, the growth rate approaches the theoretical prediction which is denoted by the solid line in the figure. The breakup time increased as the initial radius of the cylinder increased. According to the linear stability analysis, wavenumbers greater than the critical value $\tilde{k}=1$ correspond to stable modes with negative growth rates. However, as shown in Fig.~\ref{fig:RP_breakup_omega}, the growth rate remains positive even for $\tilde{k}>1$. This is likely because, due to the atomic nature of the system, unstable modes other than the imposed wavenumber grew and lead to breakup.
On the other hand, around $\tilde{k} = 0.6$, where the growth rate reaches its peak, the simulation results agree with the theory for large radius. This implies that the imposed perturbation grows faster than other perturbations arising from the thermal fluctuations of the system.

\begin{figure}[htbp]
    \centering
    \includegraphics[scale = 1]{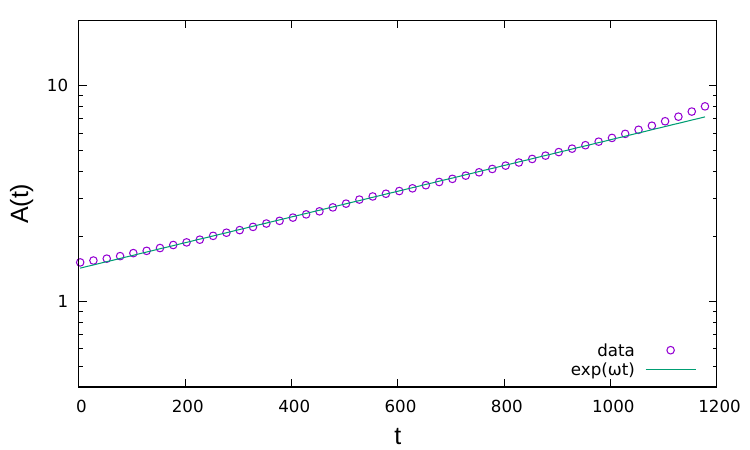}
    \caption{The time evolution of the amplitude $A(t)$ of the imposed wavenumber $\tilde{k}$. The initial radius of the cylinder is $r_0 = 15$ and the imposed wavenumber is $\tilde{k} = 0.66$, respectively. The decimal logarithm is taken for the vertical axis. The amplitude grows exponentially in the early stage.}
    \label{fig:RP_fitting}
\end{figure}

\begin{figure}[htbp]
    \centering
    \includegraphics[scale = 1]{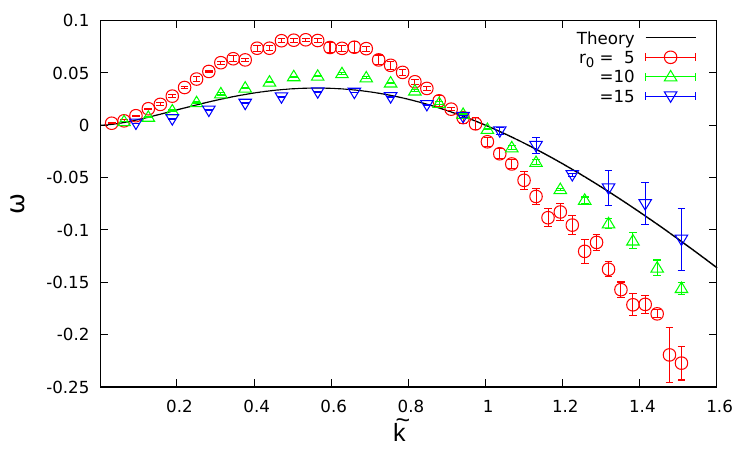}
    \caption{The dispersion relation determined from the time evolution of the amplitude with specified wavenumber. The error bars represent the average over five trials. For $r_0 = 15$, the result shows good agreement with the linear stability analysis.}
    \label{fig:RP_growth_rate}
\end{figure}

The method of determining the growth rate from the breakup time exhibits the strong finite-size effect and could not capture the negative growth rates of stable modes. Therefore, we instead determined the growth rate from the time evolution of the amplitude of the specified wavenumber.
Figure~\ref{fig:RP_fitting} shows the time evolution of the amplitude $A(t)$ of the imposed wavenumber $\tilde{k}$. Since the amplitude $A(t)$ grows exponentially, we determine the growth rate $\omega(\tilde{k})$ assuming the relation Eq.~(\ref{eq:omega_from_exp}). Repeating this procedure for various wavenumbers, we obtained the wavenumber dependence of the growth which is shown in Fig.~\ref{fig:RP_growth_rate}.

Compared to the dispersion relation obtained from the breakup time (shown in Fig.~\ref{fig:RP_breakup_omega}), it can be seen that the finite-size effects are significantly alleviated. While the dispersion relation for $r_0=5$ still deviates from the theoretical prediction, that for $r_0=15$ shows good agreement. Furthermore, negative growth rates were correctly determined in the stable region ($\tilde{k}>1$).
From the above, it was found that the method of analyzing the time evolution of the amplitude of the imposed wavenumber allows the effect of thermal fluctuations to be neglected, and as a result, the dispersion relation obtained here agrees well with the Stone's theoretical prediction even for the very small size of $r_0 = 15$.

\subsection{Unperturbed Configuration}

\begin{figure}[htbp]
    \centering
    \includegraphics[width = 0.8\textwidth]{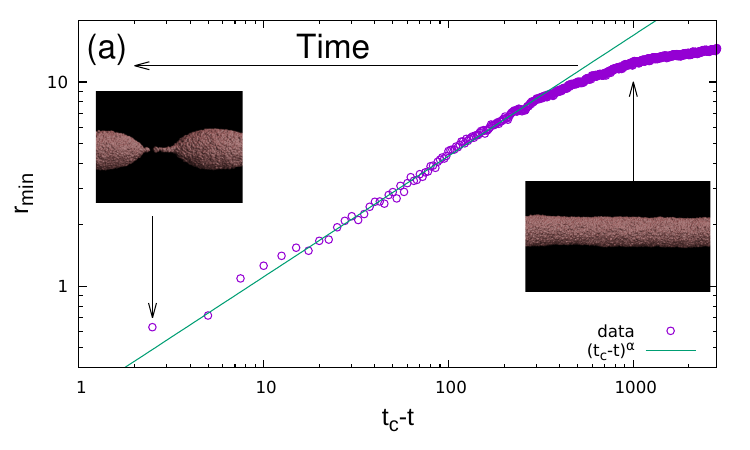}
    \includegraphics[width = 0.8\textwidth]{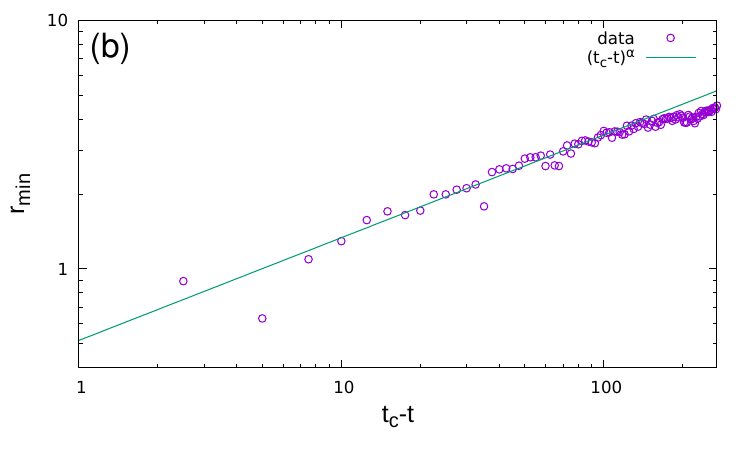}
    \caption{The time evolutions of the minimum radius $r_\mathrm{min}$ plotted against the remaining time until breakup $t_c-t$, where $t_c$ is the breakup time. The decimal logarithm is taken for both axes. (a) The case of a larger radius ($r_0 = 15$). (b) The case of a small radius ($r_0 = 5$). The solid lines denote the power-law fitting with the form $(t_c-t)^\alpha$.}
    \label{fig:tc_rmin}
\end{figure}

While we successfully reduced the effect of thermal fluctuations by imposing a perturbation on the initial condition, in this section, we performed simulations without perturbations to investigate the influence of thermal fluctuations.

We performed simulations until the unperturbed cylinder broke up and determined the breakup time $t_c$ and observed the time evolution of its minimum radius $r_\mathrm{min}$. We determined error bars from nine independent simulations for each parameter set.

Figure~\ref{fig:tc_rmin} shows the time evolution of the minimum radius $r_\mathrm{min}$ plotted against the remaining time until breakup $t_c-t$ for $r_0 = 5, 10,$ and $15$. Note that in this figure, the right side corresponds to the initial state of the system, while the left side represents the late stage. In particular, the far right of the graph corresponds to the initial state of the simulation, i.e., $t=0$.

From this figure, it can be seen that in the early stage of the system with a liquid-liquid interface, the minimum radius changes gradually. In previous research on gas-liquid systems, a rapid change was observed in the early stages of the simulation. Since the change was gradual at the liquid-liquid interface, it is suggested that the rapid change observed at the gas-liquid interface was caused by factors other than interfacial instabilities such as evaporation.

\begin{figure}[htbp]
    \centering
    \includegraphics[scale = 0.9]{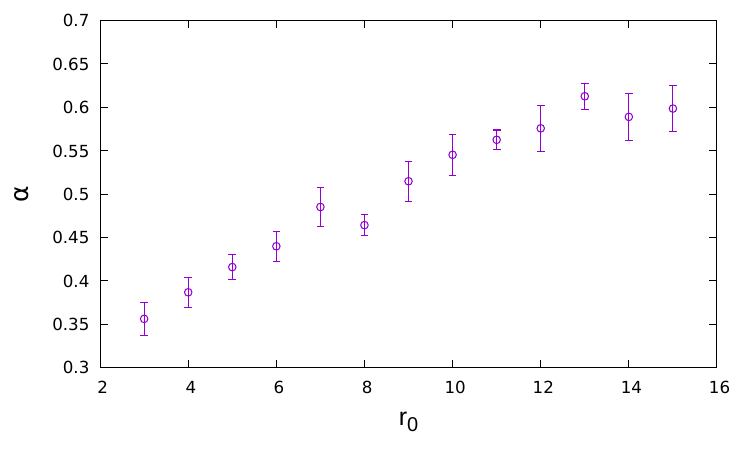}
    \caption{Dependence of the exponent $\alpha$ on the initial radius $r_0$. As the radius increases, the exponent increases.}
    \label{fig:order_param}
\end{figure}

Assuming the relation $r_\mathrm{min} \propto (t_c-t)^\alpha$, we determined the exponent $\alpha$. We evaluated this exponent for various initial radii. The initial radius dependence of the exponent is shown in Fig.~\ref{fig:order_param}. It was found that the value of the exponent increases as the initial radius increases. Since a stronger influence of thermal fluctuations is expected to result in a smaller exponent, this is consistent with the theoretical prediction\cite{eggers1993universal,eggers2002dynamics,gopan2014rayleigh}. Note that values lower than the lower bound reported by Eggers \textit{et al.} (\(\alpha \simeq 0.418\)) were observed.

From the above, for the relationship between the minimum radius and the time remaining until breakup, as the radius increases, the exponent increases and the system behavior converges to that of a macroscopic cylinder without thermal fluctuations.

\section{Discussion and Summary}\label{sec:summary}

We performed MD simulations to compare RP instability and linear stability theory. By preparing a liquid-liquid interface as the initial state, we were able to investigate instability without being affected by evaporation, where the gas-liquid system suffers from it. To the best of our knowledge, this is the first direct verification of Stone's theoretical prediction, which was derived under the condition that the physical properties of the fluid inside and outside the cylinder are identical.

We first observed the time evolutions of the interfaces from initial conditions with disturbances of specified wavenumbers. We considered two observables, the breakup time and the time evolution of the amplitude of the imposed disturbance. The growth rates estimated from the two observables approach those predicted by the classical theory as the radius of the cylinder increases. In particular, the growth rate calculated from the time evolution of the amplitude was almost perfectly consistent with the classical theory at $r_0=15$. The fact that thermal fluctuations are practically negligible and consistent with a continuum description in a system as small as $15$ atoms in diameter is remarkable, and this further suggests that particle discreteness does not strongly influence the RP instability.

We also considered the RP instability at the unperturbed initial condition. We observed the time evolution of the minimum radius $r_\mathrm{min}$. We assumed the power-law behavior $r_\mathrm{min} \propto (t_c - t)^\alpha$ where $t_c$ is the breakup time. The exponent $\alpha$ increased as the radius increased. This strong finite-size effect suggests that the impact of thermal fluctuations is prominent in this system. Nevertheless, the results with the single-mode perturbation, having a relatively small amplitude, agreed well with the classical theory, which neglects the effects of thermal fluctuations.

A remaining issue is to quantify the influence of thermal fluctuations. While this study clarified that thermal fluctuations cause deviations from classical theory, the extent of that effect remains unclear. By quantifying the effect of thermal fluctuations, the validity of hydrodynamic equations in microscopic environments may be determined. When the radius is small, the Young-Laplace pressure creates a pressure difference between the inside and outside of the cylinder. In addition, the breakup exponent obtained in the unperturbed system was found to be smaller than the value reported by Eggers \textit{et al.} (\(\alpha = 0.418\)). At present, it is not clear whether this discrepancy originates from the use of a liquid-liquid system, and further investigation is required to clarify this issue. Since Stone's theory discusses the instability at the interface of liquids with the same viscosity, the extent to which this pressure difference affects the results needs to be evaluated separately. Evaluating this effect is another important issue for future work.

\begin{acknowledgments}
    The authors would like to thank H. Noguchi and H. Nakano for fruitful discussions. This research was supported by JSPS KAKENHI, Grant No.~JP21K11923. The computati on was partly carried out using the facilities of the Supercomputer Center, Institute for Solid State Physics (ISSP), the University of Tokyo (ISSPkyodo-SC-2025
    -Cb-0038).
\end{acknowledgments}

\appendix
\section{Relationship Between Breakup Time and Growth Rate}

\begin{figure}[htbp]
    \centering
    \includegraphics[scale = 0.65]{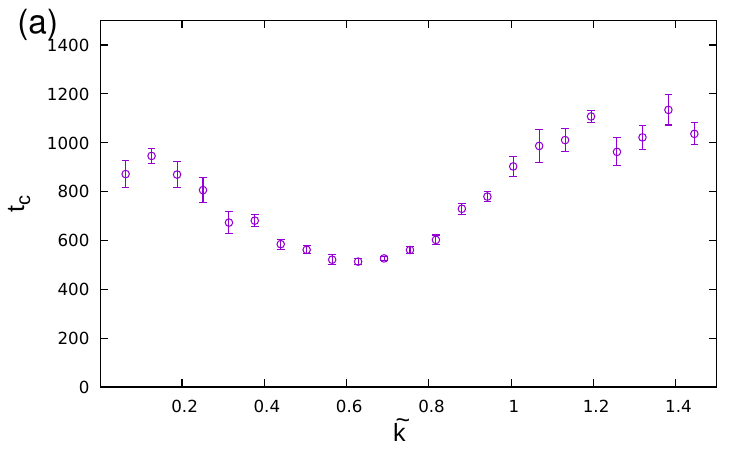}
    \includegraphics[scale = 0.65]{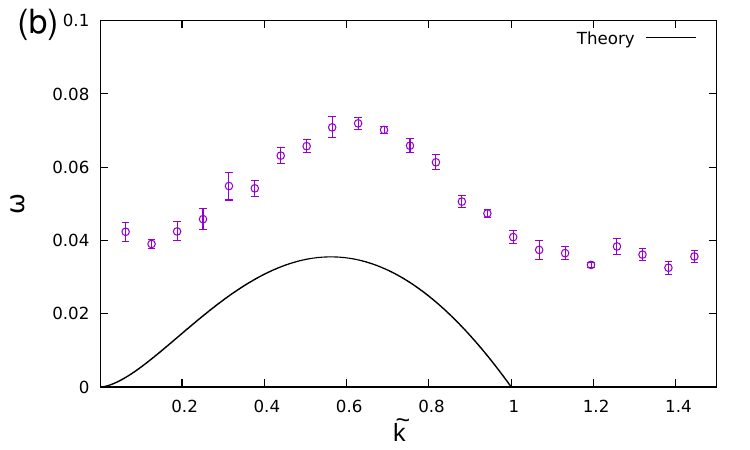}
    \caption{(a) The wavenumber dependence of the breakup time. The horizontal axis represents the wavenumber $\tilde{k}$ of the imposed perturbation in the initial state, and the vertical axis represents the time at which the breakup occurs. (b) The wavenumber dependence of the growth rate. The solid line denotes the growth rate predicted by the classical theory. The obtained growth rates take finite positive values even in the region where the theoretically predicted values are negative ($\tilde{k} \geq 1$).}
    \label{fig:RP_breakup_omega_10}
\end{figure}

This section describes the method to estimate the growth rate from the breakup time. Figure~\ref{fig:RP_breakup_omega_10}~(a) shows the dependence of the breakup time $t_c$ on the wavenumber $k$ imposed as an initial perturbation. From a obtained breakup time $t_c(k)$, we estimated a growth rate $\omega(k)$ using Eq.~(\ref{eq:RP_breakup_omega}). The estimated growth rate is shown in Fig.~\ref{fig:RP_breakup_omega_10}~(b). From the assumptions of the linear stability analysis, if the growth rate is negative, the rupture time must be infinite. However, the observed breakup time was always finite. This discrepancy is attributed to spontaneous thermal fluctuations in the system, which generated disturbances and led to instabilities. As a result, the estimated growth rate tends to be overestimated compared to the theoretical prediction.

\section{Determining Interfacial Tension and Viscosity}\label{sec:appendix_itf}

\begin{figure}[htbp]
    \centering
    \begin{picture}(10,200)
        \put(0,200){(a)}
    \end{picture}
    \includegraphics[width = 0.45\textwidth]{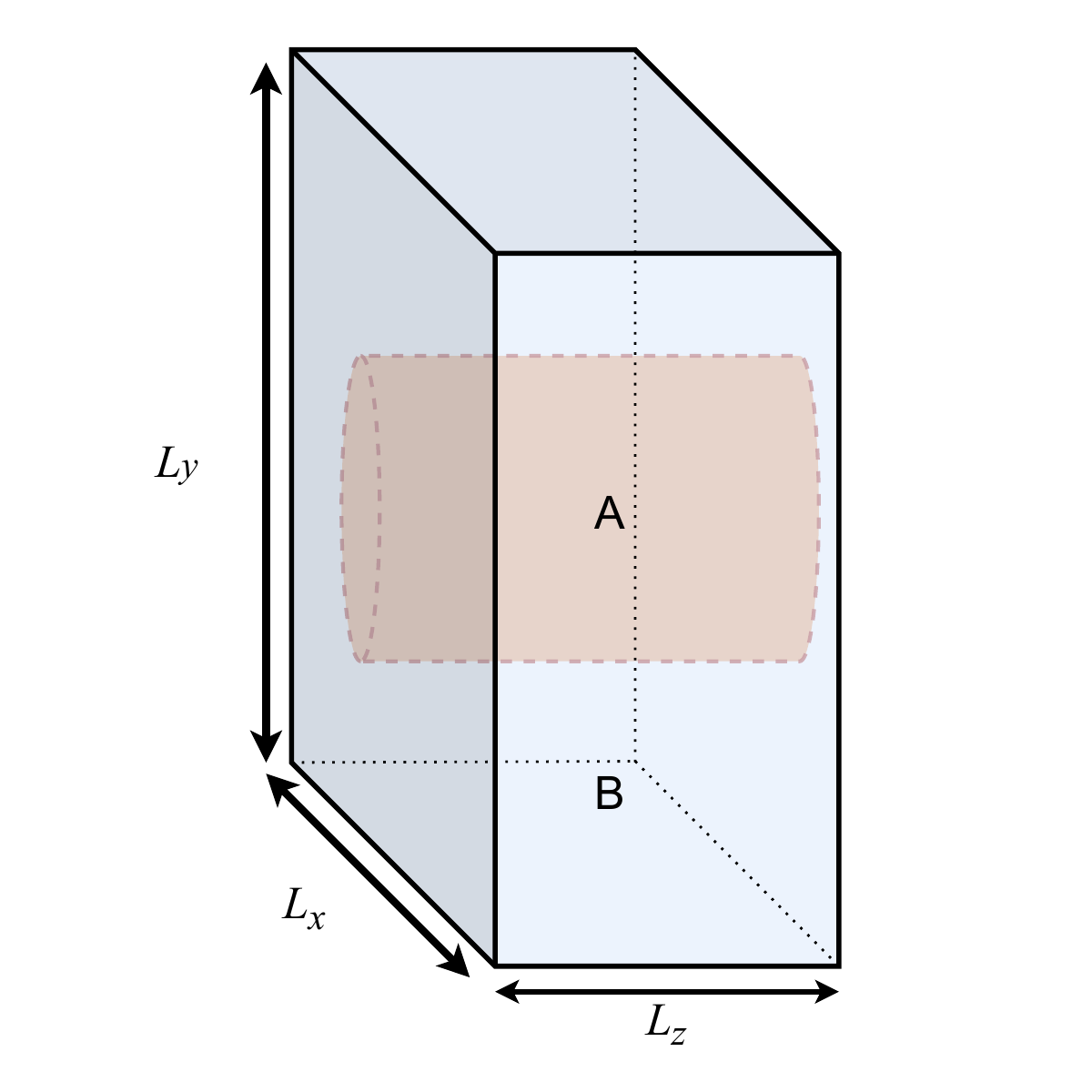}
    \begin{picture}(10,200)
        \put(0,200){(b)}
    \end{picture}
    \includegraphics[width = 0.45\textwidth]{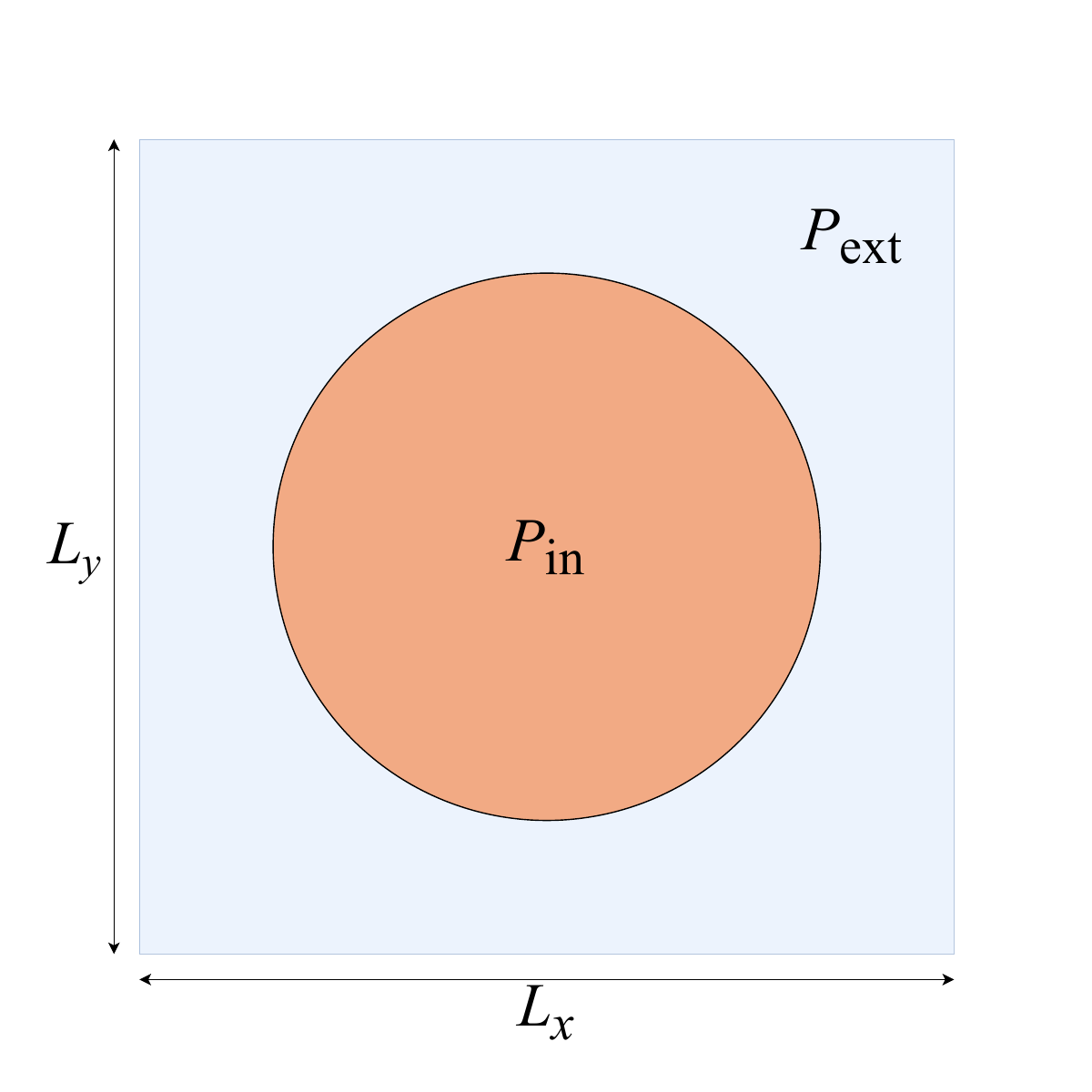}
    \caption{(a) A simulation box to determine the interfacial tension on a cylindrical surface. (b) Pressure inside and outside the cylinder.}
    \label{fig:itfForCylinder}
\end{figure}
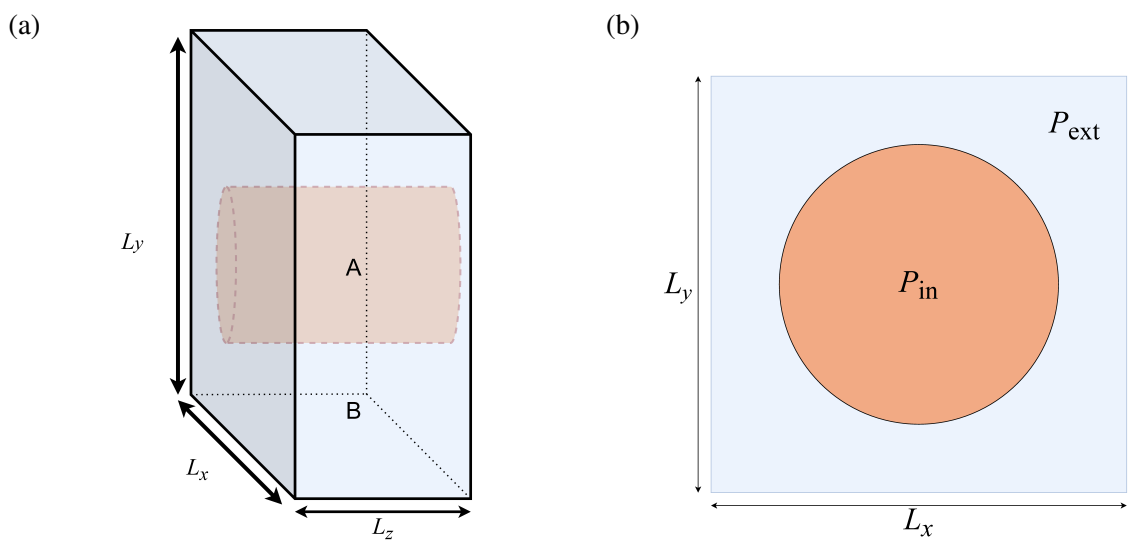

This section describes the method for determining the interfacial tension and viscosity at a cylindrical liquid-liquid interface. To determine the interfacial tension at the cylinder interface, we prepared a thin rectangular simulation box as shown in Fig.~\ref{fig:itfForCylinder}~(a). We set $(L_x,L_y,L_z) = (100,100,25)$ so that $2 \pi R \geq L_z$, where $R$ is the radius of the cylinder. In this system, the interface remains stable.

We first thermally equilibrated the single-component system, then realized the cylinder interface by changing the labels of the atoms inside the cylinder, and then further thermally equilibrated the system. Then, a pressure difference appeared between the inside and outside of the cylinder due to the interfacial tension. As shown in Fig.~\ref{fig:itfForCylinder}~(b), if the pressure inside the cylinder is $P_\mathrm{in}$ and the pressure outside is $P_\mathrm{ext}$, the following Young-Laplace equation holds.
\begin{equation}
    \Delta P = P_{\mathrm{in}}-P_{\mathrm{ext}} = \frac{\gamma}{R},
\end{equation}
where $\gamma$ is the interfacial tension on the cylindrical surface.
By measuring the pressure difference $\Delta P = P_{\mathrm{in}}-P_{\mathrm{ext}}$, we determined the interfacial tension $\gamma$.

\section{Effect of the Langevin Thermostat}\label{sec:appendix_langevin}

\begin{figure}[htbp]
    \centering
    \includegraphics[width = 0.9\textwidth]{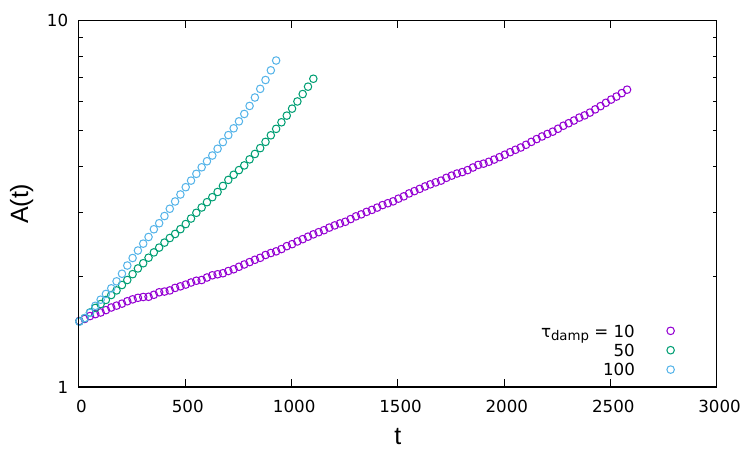}
    \caption{Relationship between the damping parameter $\tau_\mathrm{damp}$ of the Langevin thermostat and the growth dynamics of the instability amplitude. When the relaxation time is too short ($\tau_\mathrm{damp} = 10$), the breakup tends to be delayed, whereas for sufficiently large relaxation times ($\tau_\mathrm{damp} = 50, 100$), the variation in the breakup time becomes small.
    }
    \label{fig:langevinForCylinder}
\end{figure}

We used the Langevin thermostat for temperature control. In this method, friction and random forces are applied to atoms, and the temperature is regulated by the balance between these two effects. While the target temperature is determined by the ratio of the friction coefficient to the random force, the relaxation time toward the target temperature is controlled by the friction coefficient. The Langevin thermostat involves two additional forces, the frictional force $F_f$ and the random force $F_r$. The two forces are expressed as
\begin{equation}
\begin{aligned}
F_f &= - \frac{m}{\tau_\mathrm{damp}} v, \\ 
F_r &= \hat{R} \sqrt{\frac{k_B T m}{dt \tau_\mathrm{damp}}},
\end{aligned}    
\end{equation}
where $m$ is the mass of the atoms, $v$ is the velocity, $k_B$ is the Boltzmann constant, $T$ is a designed temperature, $dt$ is the time step, $\tau_\mathrm{damp}$ is the damping parameter, and $\hat{R}$ is the white noise, respectively\cite{BRUNGER1984}.
The damping parameter $\tau_\mathrm{damp}$ is the inverse of the friction coefficient and determines the relaxation time to the target temperature. When the friction coefficient is large, atomic motion is strongly suppressed, and therefore the temperature control may influence the breakup dynamics near the pinch-off event.

To examine this effect, we performed simulations with different values of the damping parameter of the Langevin thermostat. Figure~\ref{fig:langevinForCylinder} shows the relationship between the damping parameter and the temporal evolution of the perturbation amplitude. We found that when the relaxation time was too short, i.e., when the temperature control was excessively strong, the breakup dynamics became slower and the breakup time tended to increase. In contrast, when the relaxation time was increased beyond a certain value, corresponding to weaker thermostatting, the variation in breakup time became small. Although changing the relaxation time also affects the effective viscosity and thus the breakup time does not strictly converge to a unique value, we consider that the relaxation times used in our simulations were sufficiently large and that the reported dynamics are not contaminated by significant thermostat-induced artifacts.

\section{Relationship Between Temperature and the Breakup Exponent}\label{sec:appendix_temperature}
\begin{figure}[htbp]
    \centering
    \includegraphics[width = 0.9\textwidth]{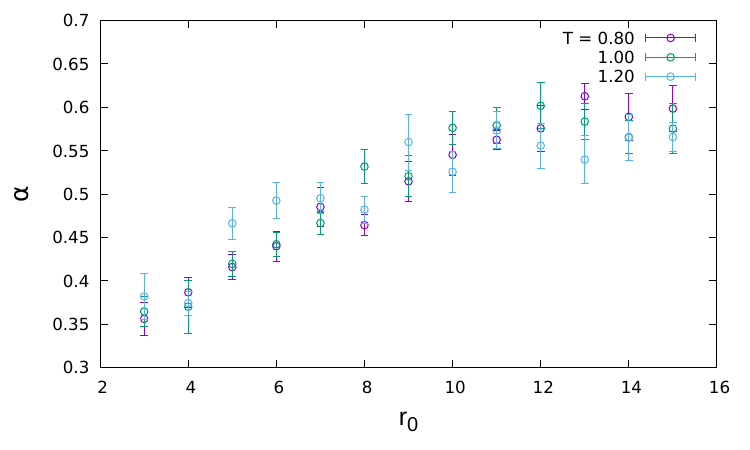}
    \caption{Relationship between the breakup exponent and the cylinder radius at different temperatures. No significant change in the exponent was observed with respect to variations in the radius.}
    \label{fig:temperatureForCylinder}
\end{figure}

The stochastic lubrication equation (SLE) proposed by Moseler \textit{et al.} includes a thermal fluctuation term in the Navier-Stokes equation, and the coefficient of the thermal noise depends on temperature~\cite{moseler2000formation}. Eggers \textit{et al.} pointed out that this thermal fluctuation term leads to a reduction in the breakup exponent. In this section, we performed simulations at different temperatures to investigate the effect of temperature on the breakup exponent. The results are shown in Fig.~\ref{fig:temperatureForCylinder}. We found that the radius dependence of the breakup exponent showed little change with temperature.

Lowering the temperature too much leads to freezing of the system. To explore a wider temperature range, it would be necessary to investigate higher-temperature regimes; however, at higher temperatures the interface becomes increasingly diffuse, which in turn requires simulations with larger system sizes.

\bibliography{reference}
\end{document}